\documentclass[aps, preprint, eqsecnum, showkeys]{revtex4}
\usepackage{graphics}
\usepackage{graphicx}
\usepackage{epsfig}
\begin{document}
\title{Superstructure in nano-crystalline $Al_{50}Cu_{28}Fe_{22}$ alloy}
\author{T.P. Yadav}
\email{yadavtp22@rediffmail.com}
\author{N.K. Mukhopadhyay$\dagger$}
\author{ R.S.Tiwari}
\author{O.N.Srivastava}
\affiliation{Department of Physics, Banaras Hindu University, Varanasi-221 005, India, \\
 $\dagger$ Department of Metallurgical Engg., IT, Banaras Hindu University Varanasi-221005, India}
\begin{abstract}
{\it This work reports the formation of nano- crystalline $Al_{50} Cu_{28} Fe_{22}$ by 
high-energy milling. 
For obtaining the nano-crystalline material, the $Al_{50} Cu_{28} Fe_{22}$ alloy 
synthesized through  slow cooling of the molten alloy was subjected to ball milling,  
which was carried out in attritor mill at $400 rpm$ for $5 h , 10 h, 20 h , 40 h$  and $80 h$ 
with a ball to powder ratio $40 : 1$ in hexane medium. The x-ray diffraction  observation of  
ball-milled samples revealed that the milling duration of $5h$ to $40 hrs$ has led to the 
formation of nano-phase. The average crystallite size comprising the nano-phase has been 
found to be $\sim 17 nm$. When the nano-crystalline alloy,  $Al_{50} Cu_{28} Fe_{22}$ was vacuum 
annealed at a temperature  of  500$^0C$ for 5 to 20 $hrs$, new structural phases 
representing  superstructures of the parent nano-crystalline  phase were found . 
The superstructure have been found to correspond to simple cubic with $a = \sqrt 2a_p$ and 
face central cubic with $a = 2a_p$  ( $a_p$ = lattce parameter of  parent nano-crystalline alloy). 
It has been proposed that the formation of different type of superstructure resulting 
due to different duration of ball milling followed by annealing is possibly governed 
by minimization of free energy of the disordered B2 phase.}
\end{abstract}

\maketitle

\section{\bf Introduction} 
Recently there has been a considerable scientific and technological interest in the 
formation of nano-crystalline/quasicrystalline phase in the $Al-Cu-Fe$ alloys by 
mechanical milling [1].  Quasicrystals have many properties which make them interesting 
for industrial applications like light weight, large strength to weight ratio and 
high hardness with a low frictional coefficient [2]. Nanostructured material, 
which can be defined as a material with the crystallite size less than $100 nm$ are 
synthesized by either bottom-up or top-down processes [3]. The bottom up approach 
starts with atoms, ions or molecules as building blocks and assembles nanoscale 
clusters or bulk material from them. The top down approach for processing of  
nanostructured materials starts with bulk solid and ends in obtaining a nano
ostructured phase through special processing routes e.g. mechanical milling, 
re-solidification through chemical methods etc. Nano-phase materials have significantly 
different behaviour from their macroscopic counterparts because their sizes are smaller 
than the characteristic length scales of physical phenomena occurring in bulk materials [4].  
The nanostructure materials are produced by using various methods, among which 
high-energy ball milling (BM) which is also known as mechanical milling (MM) 
has attracted much attention [5]. The advantages of high-energy ball milling for 
the synthesis of nano-structured materials are the formation of a more homogeneous 
product and good reproducibility [6]. The mechanical milling technique has been used 
to obtain amorphous alloys [7], high coercively permanent magnetic metallic compounds [8] 
and quasicrystals [9-10]. The formation of a quasicrystalline phase by BM/MM has been 
reported in a number of $Al$ and $Ti$ based systems [11]. Recently Mukhopadhyay et al [12] 
have studied the effect of mechanical milling on the stability of $Al$-rich, $Al-Cu-Fe$ 
and $Al-Cu-Co$ quasicrystalline alloys. They have reported that the icosahedral 
quasicrystalline phase in $Al-Cu-Fe$ system undergoes transformation to a bcc (B2 type) 
crystalline phase as a result of ball milling [13]. In this case, B2 phase does not 
transform into any other crystalline/quasicrystalline phase during isothermal 
annealing at 850$^0C$ up to 20h. It has been concluded that the B2 phase is more stable 
than the icosahedral quasicrystalline phase at those compositions.  It should be 
mentioned that $Al$-rich $Al_{65}Cu_{20}Fe_{15}$ system is of significant interest due to the 
high temperature structural stability of icosahedral quasicrystalline phase. 
The available phase equilibria data indicate that the B2 phase is a major phase 
on the $Al$-deficient side of stoichiometry [14]. The B2 structure can be understood 
in term of ordering in a bcc lattice and converting it to be a simple cubic lattice. 
Therefore, unlike a bcc lattice, one type of atom occupies the body-centered position 
and another type occupies the cube corners in the ordered lattice. When the 
composition deviates from the stoichiometry, the compositional defects must be 
introduced to preserve the crystal structure. Its unit cell contains two different 
atoms located respectively at the vertex and at the center of the cube. It is one 
of the basic simple structures that can transform into more complex structures via 
twinning at the atomic level, termed as chemical twinning [15]. The B2 type phase 
is often present together with the quasicrystals and has fixed coherent orientation 
relationship with the latter [16-17]. The detailed investigation of the B2 phase 
is also important due to its practical applications [18-19]. Though quasicrystals 
have many curious properties, they are also extremely brittle, porous and 
composition-sensitive. It is therefore interesting to substitute them by approximant 
materials, particularly B2 based ones, which are more easily prepared and have 
similar performance characteristics [20].
The purpose of the present study is to investigate the influence of high-energy 
ball milling on the phase stability, crystallite size,  lattice strain and 
lattice parameter of B2 phase formed in the pre-alloyed $Al_{50}Cu_{28}Fe_{22}$ sample. 
Present investigation clearly shows the evolution of ordered simple cubic 
phase ($a_{sc} =4.12\AA$) and as well as fcc ($a_{fcc}=5.8 \AA$) $\tau 2$ phase after milling followed 
by annealing. The evolution of the nano-structure at different stages of ball 
milling has also been investigated. 

\section{\bf Experimental Details:}  

	An alloy of composition $Al_{50}Cu_{28}Fe_{22}$ was prepared by melting the high purity 
$Al, Cu$ and $Fe$ metals in an induction furnace, in the presence of dry argon atmosphere. 
The ingot formed was re-melted several times to ensure better homogeneity. The as-cast 
ingot was crushed to particles less than $0.5$ mm in size and placed in an attritor ball 
mill (Szegvari Attritor) with  a ball to powder weight ratio of $40:1$. The attritor 
has a cylindrical stainless steel tank of inner diameter $13 cm$ and the angular 
speed of mill was maintained at 400 rpm. The milling operation was conducted from 
5 to 80h using hexane as a process control agent.  The powder obtained after 
$10h$ and $80h$ of milling were annealed isothermally at $500 ^0C$ for 5 to $20h$ in the 
evacuated quartz capsules (with vacuum of $10^{-6}$ $torr$). The milled and heat-treated 
powders were characterized by powder X-ray diffraction (XRD) using a Philips 1710 X-ray 
diffractometer with $CuK_\alpha$ radiation. The effective crystallite size and relative 
strain of mechanically milled powders as well as heat-treated products were calculated 
based on line broadening of XRD peaks. The use of the Voigt function for the 
analysis of the integral breadths of broadened X-ray diffraction line profiles 
forms the basis of a rapid and powerful single line method of crystallite-size 
and strain determination. In this case the constituent Couchy and Gaussian 
components can be obtained from the ratio of full width at half maximum 
intensity($2\omega$) and integral breadth ($\beta$) [21]. In a single line analysis the apparent 
crystallite size 'D'  and strain 'e'   can be related to Couchy ($\beta_c$)  and Gaussian 
($\beta_G$)  widths of the diffraction peak at the Bragg angle   ;
\begin{equation}
	            D = \frac{\lambda}{ \beta_c cos\theta} 	
\end{equation}

and 
\begin{equation}
                        e = \frac{\beta_G} {4 tan\theta} 
\end{equation}

The constituent  Couchy  and Gassian components  can be given as 
\begin{equation}
              \beta_c = (a_0+a_1\psi+a_2\psi^2)\beta 					 
\end{equation}

\begin{equation}
	\beta_G = [b_0 + b_{1/2} (\psi-2/\pi)^{1/2}+b_1\psi+b_2\psi^2)]\beta 
\end{equation}
where $a_0, a_1$ and $a_2$ are  Couchy constants $b_0, b_{1/2}, b_1$ and $b_2$ are Gassian constants 
and $\psi$ is $2\omega/\beta $  where $\beta$  is the integral breadth  obtained from XRD peak. The value 
of Couchy and Gassian constant have taken from the table of Langford [21] 
\vspace{0.5cm}

$a_0 = 2.0207,\hspace{1cm}     a_1 = - 0.4803, \hspace{1cm}      a_2 = - 1.7756$

$b_0 = 0.6420, \hspace{1cm}  b_{1/2} = 1.4187,\hspace{1cm}      b_1 =   2.2043,\hspace{1cm}     b_2 = 1.8706$

\vspace{0.5cm}

	From these, we have calculated the crystallite size D 
and the lattice strain 'e' for the milled $Al_{50}Cu_{28}Fe_{22}$ powders. 

\section{\bf Results and discussion}

 The X-ray diffraction (XRD) patterns for the $Al_{50}Cu_{28}Fe_{22}$ alloy obtained 
after various milling durations has been shown in figure 1. 
Curve (a) shows the B2 phase 
obtained from the as-cast ingot material and curve (b), (c), (d), and (e) 
are the XRD patterns from the powder milled for $5 h, 10h, 20h, 40h$ and $80h$ 
respectively. It can be seen from Fig.1 that the peak corresponding to the 
(110) profiles of the B2 phase becomes broader and the intensity gets reduced 
with increasing milling time. These two effects are mainly attributed to the 
increase of the internal lattice strain and reduction of the grain size. The 
evolution of the nano crystalline phase in $Al_{50}Cu_{28}Fe_{22}$ can be easily noticed 
from the increase in the broadening of x-ray diffraction lines (Fig.1) during 
different period of ball milling. It should be noted that the shift of the peaks 
towards lower $\theta$ angle side with the increase in milling time indicates the 
increase in lattice parameter. Intensity of (110) peak goes on decreasing with 
increasing milling time. After $80h$ of milling, a diffuse broad peak appears 
indicating  the transformation of the B2 phase to an amorphous phase. (see Fig. 1e).
  XRD pattern clearly indicates that the initial sharp diffraction lines 
get considerably broadened after ball milling, suggesting that the nano-crystalline 
phase appears as a result of milling.  

  To study the effect of annealing on the MM powders, the samples milled for 
10 hrs and 80hrs were subjected to annealing for various time periods. 
Fig. 2 (a), (b),(c) and (d) show the XRD pattern obtained after $10 h$ of  ball 
milling followed by annealing at $500 ^0C$ for $0h, 5h, 10h$ and $20h$ respectively. 
The XRD patterns corresponding to $10h$ and $20h$ , the  annealed samples (Fig.2 (b,c)) 
have been indexed using simple a cubic structure with $a_{sc} = 4.1\AA$ . The most 
interesting feature to be noted is that these samples show cubic structure, 
which is a superstructure of the B2 phase. The lattice parameters of the 
superstructure phase and B2 phase are interrelated as $a_{sc}=\sqrt{2} a_p$, where $a_p$ is 
the lattice parameter of the B2 phase which is a parent phase. The formation of 
the superstructure due to the ball milling and subsequent annealing has been 
observed for the first time in $Al-Cu-Fe$ alloy. We propose a structural model for 
this new superstructure of B2 phase.  Fig. 3 shows the structural model of 
superstructure with lattice parameter $\sqrt{2} a$ times of B2 phase  ($a_p =2.911\AA$), which 
is obtained from the powder after ball milling for 10 h and subsequent annealing 
at 500$^0C$ for $20h$. Figure 3(a) shows two-dimensional model of superstructure, 
which clearly indicates that the face diagonal of a cube is playing a key role 
for the formation of the superstructure of B2 phase. The three dimensional model 
(fig. 3(b)) clearly depicts the unit cell of the superstructure of the B2 phase. 
The edge of the unit cell is the diagonal of a cube face ($\sqrt{2} a_p=4.18\AA$). 
The lattice parameter of the superstructure phase, which is calculated from 
the model, exactly matches with the lattice parameter obtained from XRD of 10h 
ball milled and 20h-annealed powders.

  Fig. 4 (a) (b) (c) and (d) show the XRD patterns corresponding to $80 h$ ball 
milled $Al_{50}Cu_{28}Fe_{22}$ powders annealed at 500$^0C$ for 0, 5, 10 and $20h$ respectively. 
In the case of $20h$ annealing, the sample has been cooled slowly to avoid the 
quenching effects and detect any transformation during cooling. Unlike the sample 
milled for $10h$ and annealed at 500$^0C$ for 10/20 h, the XRD pattern shown in fig.4 (d) 
could be indexed only in terms of a fcc structure with $a=5.84\AA $. However, 
the lattice parameter of this fcc phase is also related to B2 phase as $a_{fcc}=2a_p$. 
The evolution of $\tau 2$ phases in 80 h ball milled and 20h annealed powder can easily 
be explained by the structural model shown in Fig.5. We propose two-dimensional 
geometrical model as outlined for the formation of $\tau 2$ phases from the present B2 phase. 
Figure 5(b) shows the relationship between quasicrystllaine and related crystalline ($\tau 2$) 
phases. The model is based on the concept of a  polytope  and an eight dimensional 
root lattice. Its basic atomic cluster forms a cuboctahedron with 12 vertices 
formed by intersection of three perpendicular squares with edge length of $\sqrt{2}. \sqrt{2} a$. 
This polyhedron is transformed into an icosahedron when the squares are changed 
into rectangles with edge length ratio of $\tau:1$. The properties of the eight 
dimensional root lattice give foundation to the possibility of mapping a 
quasicrystllaine structure on a crystalline structure . The proposed geometrical 
model can be applied also to the polymorphoic bcc-fcc transformation

\section{\bf Conclusions} 

On the basis of our present investigation it may be conclude that, the formation 
of the nano B2 phase starts after 5h of milling and gets completed after 40h 
of milling. Beyond 40h of milling the amorphous phase starts forming and the 
sample shows the coexistence of nano-crystalline and amorphous phases. After 80h 
of milling nano B2 phase transforms to amorphous phase completely.
The $Al_{50}Cu_{28}Fe_{22}$ samples ball milled for 10h and 80h and annealed subsequently 
at 500$^0C$ for 20h, transform to the simple cubic ( $a_{sc} = \sqrt{2} a_p$)  and the  fcc   
($a_{fcc} = 2 a_p$)  phases respectively.

{\bf Acknowledgement}

The authors would like to thank   Prof. A. R. Verma, Prof. S. Ranganathan, 
Prof. S. Lele and Prof. B.S.Murty for many stimulating discussions.  The financial 
support from Ministry of Non-Conventional Energy Sources (MNES), New Delhi, 
India is gratefully acknowledged. One of the authors (TPY) acknowledges the 
support of CSIR for award of senior research fellowship.


\begin{figure}
\caption{X-ray differaction patterns of as-cast powder (a), Ball-milled powder 
after various milling times 5h (b), 10h (c), 20h (d), 40h (e) and 80h (h). 
}
\caption{XRD patterns obtained from the powder after Ball milling for 10h (a) and
subsequent annealing at 500$^0C$ for 5h (b), 10h (c) and 20h (d).
}
\caption{The Structural model of the simple cubic superstructure with 
$a_{sc} = \sqrt{2}a_p$ of B2 phase, obtained from the powder after ball milling for 10 h 
and subsequent annealing at 500 $^0C$ for 20h}
\caption{The XRD pattern obtained from the powder after ball milling for 80 h, (a) and 
subsequent annealing at 500$^0C$ for 5 h (b), 10 h (c) and 20 h (d). The curve (d) 
indicates fcc superstructure with $a_{fcc} = 2 a_p$ of B2 phase ($a_p =2.92\AA$)
}
\caption{he Structural model of the fcc superstructure with $a_{fcc} = 2 a_p$ 
of B2 phase, obtained from the powder after ball milling for 80 h and 
subsequent annealing at 500$^0C$ for 20 h}
\end{figure}
\end{document}